%% file: main.tex
\documentclass[manuscript]{acmart}
\usepackage{graphicx} 
\usepackage{enumitem}
    \setlist{nosep, topsep=2pt}
\usepackage{booktabs}
\usepackage{tikz}
\usepackage{mathtools}

\copyrightyear{2026}
\acmYear{2026}
\setcopyright{usgovmixed}
\acmConference[PEARC '26]{Practice and Experience in Advanced Research Computing}{July 26--30, 2026}{Minneapolis, MN, USA}
\acmBooktitle{Practice and Experience in Advanced Research Computing (PEARC '26), July 26--30, 2026, Minneapolis, MN, USA}
\acmDOI{10.1145/3785462.3815801}
\acmISBN{979-8-4007-2377-3/2026/07}

\begin{document}
\title{Self-Admitted Technical Debt in Scientific Software: Prioritization, Sentiment, and Propagation Across Artifacts}
\author{Eric L. Melin}
    \affiliation{%
        \institution{Boise State University, Oak Ridge National Laboratory}
        \city{Boise}
        \state{ID}
        \country{USA}
}
\email{ericmelin@u.boisestate.edu}

\author{Nasir U. Eisty}
\affiliation{%
	   \institution{University of Tennessee}
	   \city{Knoxville}
	   \state{TN}
	   \country{USA}
}
\email{neisty@utk.edu}

\author{Gregory R. Watson}
\affiliation{%
	   \institution{Oak Ridge National Laboratory}
	   \city{Oak Ridge}
	   \state{TN}
	   \country{USA}
}
\email{watsongr@ornl.gov}

\author{Addi Malviya-Thakur}
\affiliation{
    \institution{Oak Ridge National Laboratory, University of Tennessee}
    \city{Knoxville}
    \state{TN}
    \country{USA}
}
\email{malviyaa@ornl.gov}

\date{Februray 2026}

\begin{abstract}
\input{sec_abstract}
\end{abstract}



\setcopyright{none}

\maketitle


\section{Introduction}
\input{sec_introduction}
\section{Background}
\input{sec_background}
\section{Methodology}
\input{sec_methodology}
\section{Results}
\input{sec_results}

\section{Discussion}
\input{sec_discussion}
\section{Threats to Validity}
\input{sec_threats}
\section{Conclusion}
\input{sec_conclusion}
\begin{acks}
\input{sec_acknowledgments}
\end{acks}
\bibliographystyle{ACM-Reference-Format}
\bibliography{bibliography}  
\end{document}

%% file: sec_abstract.tex
Self-admitted technical debt (SATD) impairs scientific software (SSW), yet its prioritization, sentiment, persistence, and propagation remains underexplored. Understanding how SSW developers express, and address SATD is crucial for improving SSW maintenance, and tooling.
This study investigates how SATD types and artifacts in SSW are prioritized, how sentiment relates to urgency, SATD removal and resolution rates, and the extent to which SATD propagates across artifacts.
We analyzed nine SSW repositories using a SATD classification model and a semantic embedding-based prioritization heuristic. SATD was examined across multiple artifacts, with sentiment assessed via a fine-tuned transformer. Propagation was traced, priority scores compared to static analysis, and removal and resolution rates quantified.
SATD in comments, commits, and pull requests receive higher priority than SATD in issues, with negative sentiment amplifying urgency. Resolution and removal rates lag behind open-source software (OSS) averages. Most SATD remains confined to the originating artifact, but longer propagation chains are rare and correlate with higher priority, highlighting persistent and high impact debt.
Prioritization is influenced by artifact type and sentiment, while low removal and resolution rates signal persistent debt. Cross-artifact propagation marks high priority, unresolved SATD, providing empirical guidance for targeted monitoring, review prioritization, and tool supported maintenance in SSW.

%% file: sec_introduction.tex
\vspace{-1mm}
Modern scientific discovery often relies on complex, long-lived software systems to support data analysis, simulation, and experimentation~\cite{johanson2018software}. The correctness, maintainability, and evolution of scientific software (SSW) are critical, as software defects or design limitations may directly affect the validity and reproducibility of scientific results~\cite{nguyen2010survey, sanders2008development}. 

Technical debt (TD) is a well established metaphor~\cite{cunningham_wycash_nodate} used to describe suboptimal design or implementation decisions that provide short-term benefits at the expense of long-term maintainability and quality. When left unmanaged, TD tends to accumulate over time~\cite{martini2015investigating}, increasing maintenance costs and reducing a system's ability to evolve. As a result, developers must continuously decide which instances of TD to address, defer, or ignore, making prioritization an essential aspect of effective debt management.
Self-admitted technical debt (SATD)~\cite{potdar2014exploratory} represents a specific subset of TD in which developers explicitly acknowledge technical shortcomings generally documented in code comments, commit messages, pull requests, or issue tracker entries within software~\cite{li_identifying_2022}. Prior research has leveraged SATD to study the nature, evolution, and resolution of TD, primarily within open-source software (OSS) ecosystems ~\cite{maldonado2017empirical, bavota2016large}.

However, little is known about how SATD manifests and is prioritized within SSW. SSW is often developed under unique constraints, including rapidly evolving scientific knowledge, domain expertise, and scientific validity constraints,~\cite{kelly2013industrial, nguyen2010survey, carver2007software}. These pressures may influence not only the accumulation of SATD~\cite{melin2026multiartifactanalysisselfadmittedtechnical}, but also how SATD is prioritized, and resolved.
To address this gap, this paper investigates the prioritization, sentiment, persistence, and propagation of SATD across artifacts in SSW. 
All figures and code can be found in the following replication package\footnote{https://github.com/MelinHead225/PEARC26-Replication-Package}.

%% file: sec_background.tex
\vspace{-1mm}
\paragraph{SATD Taxonomy \& Prioritization}
SATD refers to cases where developers explicitly acknowledge the existence of TD in code comments, documentation, or other software artifacts. SATD was first formalized by Potdar and Shihab~\cite{potdar2014exploratory}, and subsequently Maldonado and Shihab~\cite{Maldonado2015} introduced an initial classification consisting of five SATD types. 
Recent SATD research has converged on a simplified taxonomy emphasizing four primary types~\cite{pham2025descriptor, sutoyo2024satdaug, li_identifying_2022}: \textit{code/design debt}—suboptimal or temporary implementation and design choices that degrade code quality or structure; \textit{documentation debt}—incomplete or insufficient documentation supporting software artifacts; \textit{test debt}—gaps in existing test coverage or the need to enhance tests; and \textit{requirement debt}—expressions of incompleteness in methods, classes, or overall functionality~\cite{alves2014towards, Maldonado2015}. Because of 
substantial overlap in 
impact on maintenance, code debt and design debt are frequently treated as a single \emph{code/design debt} category~\cite{sutoyo2024satdaug, li_identifying_2022}. This consolidated taxonomy provides a structured basis for empirical studies and automated detection, enabling 
effective identification, prioritization, and management of SATD.

Repaying all SATD present in a system is often unfeasible due to limited development resources~\cite{alfayez2020systematic}. 
Prior research has emphasized decision models for TD prioritization, where items are ordered based on stakeholder interests, business objectives, and organizational constraints~\cite{seaman2012using, LI2014183}.
Systematic reviews further show that prioritization is often applied through categorical labels (e.g., low, medium, high), ordered rankings, or binary decisions
~\cite{alfayez2020systematic, pina2021technical}.

Building on these foundations, SATD research has explored signals to approximate prioritization. Several studies analyze temporal characteristics, showing that SATD exhibits highly variable lifespans, with some instances intentionally resolved within short release cycles (short-term SATD), while others persist for long periods~\cite{potdar2014exploratory, maldonado2017empirical}. Recent work has leveraged version-based datasets and machine learning models to predict whether SATD will be resolved before the next release, highlighting developer workload and contextual features as important predictors~\cite{huang2025empirical}. Other approaches rely on textual cues, such as priority related keywords or severity descriptors 
to rank SATD instances
~\cite{mensah2018value, de2022toward}.

Additional studies indicate that sentiment expressed in SATD comments may also influence prioritization decisions, with negatively worded SATD being substantially more likely to be addressed~\cite{cassee2025negativity}. Complementary research has focused on estimating SATD repayment effort using textual and code change features, demonstrating that different SATD types demand substantially different levels of effort~\cite{li2023presti}. However, the absence of ground truth for SATD priority and effort remains a key challenge for evaluation.

Despite these advances, most SATD prioritization studies remain OSS-centric and primarily focus on source code comments, with limited consideration of multiple development artifacts. In addition, recent evidence suggests that automated approaches, including large language models, are currently only able to resolve a small fraction of SATD automatically, further underscoring the need for effective prioritization strategies~\cite{mastropaolo2023towards}. 

\vspace{-1mm}
\paragraph{SATD in Scientific Software} In recent years, the study of SATD in SSW has received growing attention.
A preliminary study by Awon~\cite{awon2024self} manually annotated 28,680 code comments across nine SSW repositories to investigate the prevalence and characteristics of SATD in this domain.
During the annotation process, the author identified a substantial subset of comments that did not align with existing SATD categories.
Based on these observations, Awon et al. introduced a novel SATD category, \textbf{scientific debt}, defined as \textit{the accumulation of sub-optimal scientific practices, assumptions, and inaccuracies within scientific software that potentially compromise the validity, accuracy, and reliability of scientific results}.

Building on the dataset introduced by Awon et al.~\cite{awon2024self}, Melin et al.~\cite{melin2025exploring} conducted the first large-scale empirical study contrasting SATD in OSS and SSW.
Their analysis examined SATD expressed in \textit{code comments} across 27 repositories and revealed that on average SSW contains 4.93$\times$ more SATD overall than OSS, and 9.25$\times$ more instances of \textit{scientific debt}.
These findings demonstrate a substantial disparity in SATD prevalence and emphasize the need for further research into SATD management in SSW.
While the initial dataset~\cite{awon2024self} focused exclusively on code comments (CC), recent SATD research has expanded beyond single artifacts to include commit messages (CM), issue trackers (IS) sections, and pull request (PR) discussions~\cite{li_identifying_2022, xavier2020beyond}.
Melin et al.~\cite{melin2026multiartifactanalysisselfadmittedtechnical} extended Awon’s dataset to support multi-artifact SATD analysis and conducted a practitioner validation study to assess the usefulness and recognizability of \textit{scientific debt} in practice.
In their study they observed that existing models trained on traditional SATD often miss \textit{scientific debt} in SSW, emphasizing the need for its explicit detection in SSW.
Their work resulted in a manually annotated multi-artifact SATD dataset catered to SSW, and a fine-tuned SATD classification model.
Adjacent work by Sharma et al.~\cite{sharma_self-admitted_2022} investigated SATD in SSW coded in dynamically-typed languages such as R noticing that SSW differs in paradigm,
and 
semantics 
when compared to traditional OSS. They note that many software
engineering topics are understudied in SSW development, with SATD detection remaining a challenge for this domain. 

%% file: sec_methodology.tex
\vspace{-1mm}

\subsection{SSW Project Selection}
\vspace{-1mm}
Due to this study's focus on SATD prioritization, sentiment, and propagation rather than classification, we selected nine U.S. Department of Energy (DOE) Consortium for the Advancement of Scientific Software (CASS)\footnote{https://cass.community/software/} projects that were manually labeled from a previous SSW SATD study~\cite{melin2026multiartifactanalysisselfadmittedtechnical} as case studies.
For comparison with existing OSS SATD studies, it was important to ensure data availability, sufficient project activity, and exclusion of smaller projects, so the following constraints on project selection were imposed:
publicly available on GitHub, contributed to within the last four months, at least 10k commits, at least 20 contributors, at least two years old, and at least 40 stars.
\vspace{-1mm}

\subsection{Artifact Extraction \& Cross-Artifact Linkage}
\vspace{-1mm}
To investigate SATD persistence and propagation across artifacts, we constructed a per repository scoped linkage graph connecting code comments, commits, pull request sections, and issue sections. 
Code comments were obtained from a fresh local clone of each repository to preserve file-level and revision-level context. Commits, pull requests, and issues were extracted via the GitHub API. Pull requests and issues were further decomposed into fine-grained textual sections (e.g., title, description, comments), which were treated as distinct analytical units for SATD identification.

Artifact linkage was centered on \emph{commit identifiers}, which served as the primary key connecting all artifact types. 
Code comments were linked to commits using commit SHAs indicating when a comment was introduced, modified, or removed. 
Commits were linked to pull requests using repository metadata, including merge commit SHAs and explicit commit inclusion within pull requests. 
All links were constructed bidirectionally, which enabled traversal from commits to pull requests and vice versa.
Pull requests were linked to issues through explicit references, such as issue identifiers mentioned in pull request descriptions or metadata defined associations provided by GitHub. 

While linkage was constructed using main artifacts (commits, pull requests, and issues), SATD identification was performed at the section level (PR's or issue's title/description/comments). To reconcile this difference in granularity, linkage information from main artifacts was propagated to their associated sections. 
Section level artifacts inherited links from their parent artifacts. 
Using the resulting linkage graph, SATD propagation was evaluated along ordered artifact paths reflecting the typical progression of development artifacts:
\vspace{-2mm}
\[
    \textit{issue} \leftrightarrow \textit{pull request} \leftrightarrow \textit{commit} \leftrightarrow \textit{comment} 
\]
Not all artifacts were required to be present for a valid path. For example, a SATD comment may link directly to commits without an associated pull request, or a pull request may not reference an issue. Propagation depth was therefore defined by the longest ordered path that could be constructed for a given SATD artifact.
Propagation was conditioned on the presence of SATD in intermediate artifacts. Specifically, a SATD comment was considered to propagate across artifacts only if it could be structurally traced through linked commits, pull request sections, and issue sections that themselves contain SATD. While this approach does not guarantee that the same underlying SATD concern is expressed verbatim across artifacts, it captured the 
persistence of SATD throughout the development process.
The resulting artifact graph serves as the basis for downstream analyses of SATD prioritization, sentiment, persistence, and propagation.
\vspace{-1mm}

\subsection{SATD Identification \& Classification}
\vspace{-1mm}
We created a dataset of SSW SATD to analyze utilizing the SATD identification and classification model
\footnote{https://huggingface.co/MelinHead225/multitask-falcon-satd2}
fine-tuned for SSW and multi-artifact analysis from Melin et al.~\cite{melin2026multiartifactanalysisselfadmittedtechnical}.
We reused the trained model and weights without further fine-tuning for inference to rapidly construct a large set of SATD from SSW based on previous literature for our analysis. 
The combination of the nine selected SSW repositories and the analysis capabilities of the selected identification and classification model resulted in the following SATD set displayed in Table~\ref{tab:satd_artifact_counts} to be analyzed in this study.
Non-SATD instances were included to contextualize class imbalance but are excluded from downstream SATD analyses.
\vspace{-1mm}

\setlength{\textfloatsep}{6pt plus 2pt minus 2pt}
\begin{table}
\captionsetup{skip=2pt}
\centering
\footnotesize
\setlength{\tabcolsep}{4pt}
\renewcommand{\arraystretch}{0.8} 

\begin{minipage}[c]{0.58\textwidth}
\centering
\caption{Class Counts by Artifact Type}
\label{tab:satd_artifact_counts}
\begin{tabular}{lcccc|c}
\toprule
\textbf{SATD Class} & \textbf{CC} & \textbf{CM} & \textbf{IS Sec.} & \textbf{PR Sec.} & \textbf{Total} \\
\midrule
Code/Design Debt     & 12,948    & 4,024   & 3,433   & 16,254  & 36,659    \\
Documentation Debt   & 203       & 5,725   & 1,731   & 7,043   & 14,702    \\
Test Debt            & 1,371     & 6,709   & 1,012   & 5,517   & 14,609    \\
Requirement Debt     & 2,902     & 1,605   & 657     & 1,163   & 6,327     \\
Scientific Debt      & 1,529     & 294     & 392     & 228     & 2,443     \\
Non-SATD             & 1,170,677 & 258,867 & 71,956  & 355,069 & 1,856,569 \\
\midrule
\textbf{Total}       & 1,189,630 & 276,224 & 79,181  & 386,274 & 1,939,309 \\
\bottomrule
\end{tabular}
\end{minipage}
\hfill
\begin{minipage}[c]{0.38\textwidth}
\centering
\caption{Sentiment Model Performance Summary}
\label{tab:sentiment-classification}
\begin{tabular}{lcccc}
\toprule
\textbf{Class} & \textbf{Prec.} & \textbf{Rec.} & \textbf{F1} & \textbf{Supp.} \\
\midrule
non-negative & 0.85 & 0.94 & 0.89 & 72 \\
negative     & 0.82 & 0.60 & 0.69 & 30 \\
\midrule
\textbf{Accuracy}      &     &     & 0.84 & 102 \\
\textbf{Macro Avg}     & 0.83 & 0.77 & 0.79 & 102 \\
\textbf{Weighted Avg}  & 0.84 & 0.84 & 0.84 & 102 \\
\bottomrule
\end{tabular}
\end{minipage}

\end{table}

\vspace{-3mm}

\subsection{SATD Prioritization}
\vspace{-1mm}
To systematically prioritize SATD, we implemented a heuristic that assigns a continuous priority score to each SATD artifact based on textual content. While prioritization can be conceptually based on several  factors, such as debt type, estimated time to resolve, code complexity, urgency, or impact on correctness, we relied on semantic similarity to an existing curated set of SATD priority indicating terms in literature~\cite{de2022toward}.
All extracted SATD artifacts were encoded using a sentence embedding model (\texttt{all-mpnet-base-v2}). The priority score for each SATD artifact was computed as the cosine similarity between its embedding and the centroid embedding of the priority term set, with higher scores reflecting stronger alignment with high-priority indicators.
This approach is both replicable and extensible. While this study utilizes an established priority term list, the methodology could be readily adapted by updating the priority terms to reflect different developer goals or organizational priorities without altering the computational framework.
\vspace{-3mm}
\subsubsection{Priority Heuristic Ablation and Selection}
Because SATD priority is a constructed heuristic rather than a ground-truth label,
we evaluated several candidate heuristics to guide the selection of a stable prioritization metric based on the priority-indicating keyword list proposed in prior work~\cite{de2022toward}. We compared lexical overlap, TF–IDF similarity, sentence embedding similarity, and hybrid combinations thereof. For each heuristic, SATD artifacts were ranked by priority score and compared using Spearman rank correlation and Jaccard similarity top-25 overlap measures.

Embedding-based similarity consistently preserved the global ranking structure observed in more complex hybrid formulations while avoiding additional feature coupling. Simpler lexical and TF–IDF approaches surfaced different top-ranked artifacts but showed lower agreement under strict cutoffs. Based on these results, we adopt the embedding-only heuristic as the minimal formulation that maintains ranking stability while remaining extensible and replicable.
\vspace{-3mm}
\subsubsection{Priority Heuristic Comparison}
To provide an external point of comparison for our SATD prioritization heuristic, we compared SATD comments with issues detected by SonarQube~\cite{sonarqube} across the nine selected SSW repositories. 
Rantala et al.~\cite{rantala2024keyword} found that 36\% of Keyword-labeled-SATD comments are in the context of a SonarQube issues indicating that many SATD indicators are outside the detection capabilities of rule-based static analysis alone. As a result, SonarQube is not treated as a ground truth oracle for SATD, but rather as a complementary reference point reflecting established notions of code issue severity.
Each repository was analyzed using a local SonarQube server (Community Edition v9.9.8, build 100196) with the SonarScanner CLI version 8.0.1.6346. SATD instances were matched to corresponding SonarQube issues based on repository, file path, and line number proximity (within $\pm$2 lines).

\begin{figure}
\centering
\begin{minipage}[t]{0.3\textwidth}
    \centering
    \includegraphics[width=\columnwidth]{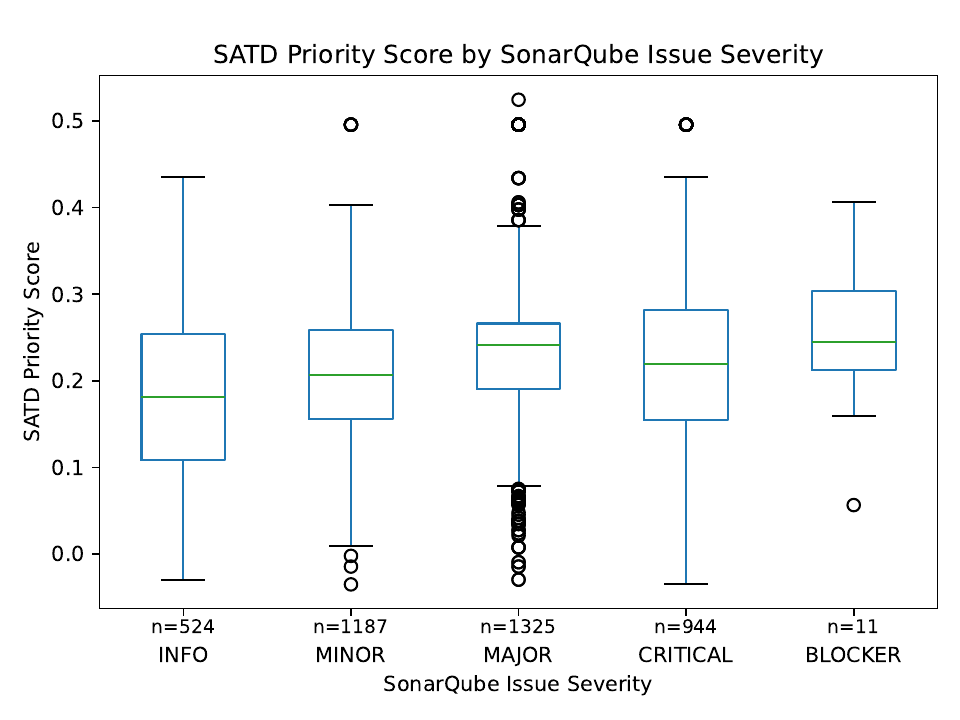}
    \caption{Distribution of SATD priority scores stratified by SonarQube issue severity.}
    \label{fig:satd_priority_severity}
\end{minipage}
\hfill
\begin{minipage}[t]{0.31\textwidth}
    \centering
    \includegraphics[width=\linewidth]{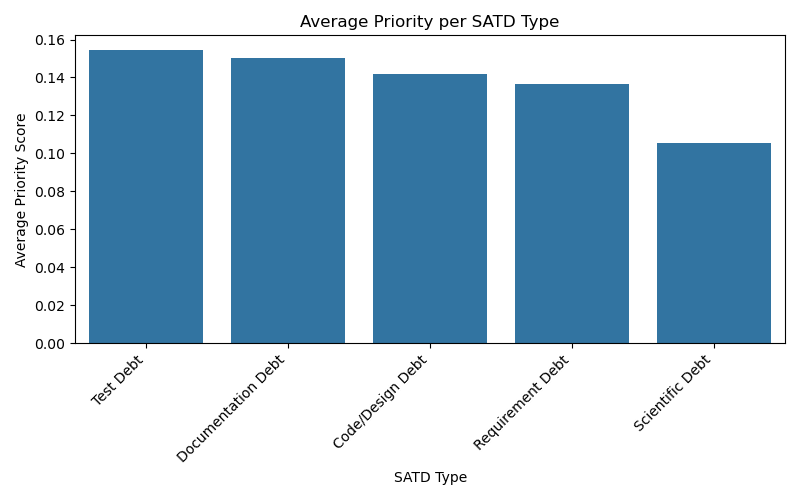}
    \caption{Average Priority per SATD Type}
    \label{fig:priority_per_satd}
\end{minipage}
\hfill
\begin{minipage}[t]{0.31\textwidth}
    \centering
    \includegraphics[width=.8\linewidth]{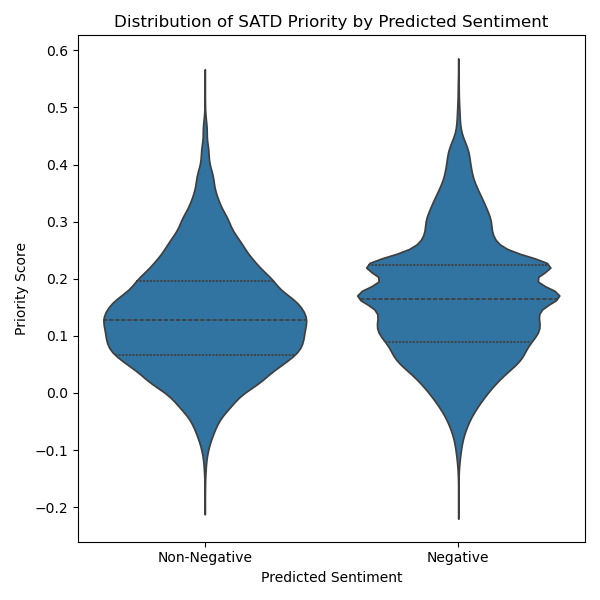}
    \caption{SATD Priority Distribution vs Sentiment}
    \label{fig:priority_vs_sentiment}
\end{minipage}

\end{figure}

Correlation analysis revealed a modest positive relationship between SATD priority scores and SonarQube issue severity (Pearson $r = 0.160$, $p < 0.001$; Spearman $\rho = 0.149$, $p < 0.001$), indicating higher priority SATD items tend to coincide with more severe static analysis findings. While strong correlations are not expected given the conceptual differences between SATD and rule based static analysis, the observed trend suggests partial alignment between developer expressed urgency and tool detected issue severity.
Figure~\ref{fig:satd_priority_severity} visualized this relationship, showing the distribution of SATD priority scores stratified by SonarQube issue severity, with the number of matched instances ($n$) indicated below each label. Median SATD scores generally increased with issue severity, supporting the use of SonarQube severity levels as a coarse external reference for validating prioritization behavior.
\vspace{-2mm}

\subsection{Sentiment Analysis}
\vspace{-1mm}
To analyze tone in SATD artifacts, we trained a transformer-based sentiment classifier to identify whether expressions convey negative sentiment. Sentiment was modeled as a binary attribute (\emph{negative} vs.\ \emph{non-negative}) to emphasize polarity and support downstream analyses.
We first evaluated the pre-trained \emph{cardiffnlp/twitter-roberta-base-sentiment-latest} model~\cite{camacho-collados-etal-2022-tweetnlp} without domain adaptation. Applied directly to SATD artifacts, it performed poorly (test accuracy 0.37) and exhibited severe class imbalance, indicating a domain mismatch between social media and technical software text. We therefore fine-tuned the model on a manually annotated dataset of 1,038 SATD instances~\cite{cassee2022self}, labeled as \emph{negative} or \emph{non-negative}, using an 80/10/10 stratified train/validation/test split.
Fine-tuning employed AdamW~\cite{loshchilov2019decoupledweightdecayregularization} with a linear learning rate schedule, gradient clipping, and early stopping (patience = 2). The adapted model achieved substantially improved performance on the held-out test set (Table~\ref{tab:sentiment-classification}), with an accuracy of 0.84 and macro F1-score of 0.79. Although performance on negative instances remained lower, the model was sufficiently reliable for comparative and exploratory analyses.
For inference, all SATD artifacts from the nine SSW repositories were tokenized and processed using the fine-tuned model to obtain predicted sentiment labels.
\vspace{-2mm}

%% file: sec_results.tex
\vspace{-1mm}

\subsection{RQ1: How do different SATD types and artifacts compare in their prioritization in SSW repositories?}
\vspace{-1mm}
Table~\ref{tab:satd_priority_artifact_type} summarizes the distribution of priority scores across SATD artifact types. 
Overall, SATD appearing in \emph{commit messages} exhibits the highest average and median priority (mean = 0.1310, median = 0.1302), followed closely by \emph{comments} and \emph{pull request sections}.
In contrast, \emph{issue sections} show the lowest prioritization (mean = 0.0862, median = 0.0830). 
This suggests that SATD identified closer to code evolution and review activities is generally perceived as more urgent than SATD documented in issue discussions.

\setlength{\textfloatsep}{6pt plus 2pt minus 2pt}
\begin{table}
\captionsetup{skip=0pt}
\centering
\footnotesize
\setlength{\tabcolsep}{2pt}

\begin{minipage}[c]{0.41\textwidth} 
\centering
\caption{Priority by artifact type for SATD artifacts.}
\label{tab:satd_priority_artifact_type}
\begin{tabular}{lrrc}
\toprule
\textbf{Artifact Type} & \textbf{N} & \textbf{Mean} & \textbf{Median} \\
\midrule
Comment       & 18,953 & 0.1204 & 0.1139 \\
Commit        & 18,357 & 0.1310 & 0.1302 \\
PRSection     & 29,852 & 0.1136 & 0.1113 \\
IssueSection  & 7,155  & 0.0862 & 0.0830 \\
\bottomrule
\end{tabular}
\end{minipage}
\hfill
\renewcommand{\arraystretch}{0.8}  
\begin{minipage}[c]{0.58\textwidth} 
\centering
\caption{SATD comment introduction, removal, and resolution rates across SSW repositories.}
\label{tab:satd-resolution-rate}
\begin{tabular}{lrrrr}
\toprule
\textbf{Repository} & \textbf{KLOC/Contr.} & \textbf{Introduced} & \textbf{Removed} & \textbf{Resolution} \\
\midrule
ADIOS2   & 19.18 &  526   & 147  & 0.279 \\
Trilinos & 73.28 & 12,871 & 4,760 & 0.370 \\
dyninst  & 12.82 &  282   & 128  & 0.454 \\
hypre    & 22.75 &  1,196 & 416  & 0.348 \\
kokkos   &  2.83 &  1,117 & 595  & 0.533 \\
legion   & 17.18 &  435   & 172  & 0.395 \\
spack    &  0.28 &  934   & 547  & 0.586 \\
tau2     & 79.96 &  1,044 & 299  & 0.286 \\
visit    & 67.23 &  548   & 82   & 0.150 \\
\midrule
\textbf{Mean}   & \textbf{32.83} &  &  & \textbf{0.378} \\
\textbf{Median} & \textbf{19.18} &  &  & \textbf{0.370} \\
\bottomrule
\end{tabular}
\end{minipage}

\end{table}
When analyzed by SATD type, we observe clear differences in prioritization as displayed in Figure ~\ref{fig:priority_per_satd}. 
\emph{Test Debt} and \emph{Documentation Debt} receive the highest average priority scores (0.1545 and 0.1502, respectively), indicating that deficiencies affecting testing  and documentation clarity are commonly perceived as requiring prompt attention. 
\emph{Code/Design Debt} and \emph{Requirement Debt} occupy a middle range, while \emph{Scientific Debt} exhibits the lowest average priority (0.1057). 
This lower prioritization may reflect the nature of the priority keywords list established in literature~\cite{de2022toward} being catered towards OSS and not SSW. 

Finally, we observe that prioritization is strongly associated with predicted sentiment as displayed in Figure~\ref{fig:priority_vs_sentiment}. 
SATD instances classified as having \emph{negative sentiment} show substantially higher average priority (0.1646) than those with \emph{non-negative sentiment} (0.1349). 
This aligns with findings from previous OSS SATD sentiment analyses~\cite{cassee2025negativity, cassee2022self}, and the intuition that emotionally charged or critical language is often used when developers perceive an issue as urgent or problematic, reinforcing sentiment as a useful signal for SATD prioritization.
\vspace{-2mm}
\subsection{RQ2: How likely are SATD comments to be removed in SSW, and how long do they persist?}
\vspace{-1mm}
To assess the lifecycle of SATD in SSW, we examine both the likelihood that SATD \textit{code comments} are eventually removed and the time elapsed between their introduction and removal.
Following prior work on SATD evolution~\cite{maldonado2017empirical}, we measure (i) the SATD resolution rate, defined as the proportion of introduced SATD code comment instances that are subsequently removed, and (ii) the removal time, measured as the number of days between SATD introduction and its 
removal. To support comparison with 
OSS studies, we also report thousands of lines of code (KLOC) per contributor.
Table~\ref{tab:satd-resolution-rate} summarizes SATD introduction and removal statistics across the studied SSW repositories. 
On average, only 37.8\% of SATD instances are eventually removed, indicating that the majority of SATD introduced into SSW persists indefinitely. 
Resolution rates vary substantially across repositories, ranging from a low of 15.0\% in \textit{visit} to a high of 58.6\% in \textit{spack} suggesting considerable differences in SATD management practices across projects.
Table~\ref{tab:satd-removal} reports the time required to remove SATD once introduced. 
For this analysis we only analyze SATD comments with eventual removal dates. 
Across repositories we observe that SATD persists for an average of approximately 690 days (nearly two years), with a median removal time exceeding 278 days. 
Several repositories exhibit particularly long persistence, with \textit{dyninst}, and \textit{visit} showing median removal times exceeding 490 days. 
These findings indicate that even when SATD is eventually addressed, remediation typically occurs only after long periods of persistence.

\setlength{\textfloatsep}{6pt plus 2pt minus 2pt}
\begin{table}
\captionsetup{skip=0pt}
\centering
\footnotesize
\setlength{\tabcolsep}{4pt}
\renewcommand{\arraystretch}{0.8} 

\begin{minipage}[c]{0.47\textwidth} 
\centering
\caption{SATD comment removal time across SSW repositories.}
\label{tab:satd-removal}
\begin{tabular}{lrrrr}
\toprule
\textbf{Repository} & \textbf{KLOC/Contributor} & \textbf{Avg. Days} & \textbf{Median Days} \\
\midrule
ADIOS2   & 19.18 & 530.9 & 216.0  \\
Trilinos & 73.28 & 521.5 & 166.0 \\
dyninst  & 12.82 & 1,198.7 & 492.0  \\
hypre    & 22.75 & 626.5 & 323.0  \\
kokkos   & 2.83  & 363.4 & 112.0  \\
legion   & 17.18 & 645.6 & 200.0  \\
spack    & 0.28  & 630.4 & 370.0  \\
tau2     & 79.96 & 464.4 & 134.0  \\
visit    & 67.23 & 1,232.2 & 496.5 \\
\midrule
\textbf{Mean}   & \textbf{32.83} & \textbf{690.4} & \textbf{278.8} \\
\textbf{Median} & \textbf{19.18} & \textbf{626.5} & \textbf{216.0} \\
\bottomrule
\end{tabular}
\end{minipage}
\hfill
\begin{minipage}[c]{0.50\textwidth} 
\centering
\caption{Propagation and priority characteristics by artifact chain length.}
\label{tab:satd_chain_length_combined}
\begin{tabular}{p{0.6cm}p{1.0cm}p{1.5cm}p{1.4cm}p{1.6cm}}
\toprule
\textbf{Chain Length} &
\textbf{Unique Chains} &
\textbf{SATD-Only Chains (\%)} &
\textbf{Mean SATD Priority} &
\textbf{Median SATD Priority} \\
\midrule
1 & 1,523,370  & 58,825 (3.86\%)  & 0.1262 & 0.1249 \\
2 & 20,450,593 & 33,113 (0.16\%)  & 0.1273 & 0.1261 \\
3 & 2,974,419  & 4,524 (0.1452\%)  & 0.1452 & 0.1461 \\
4 & 658,212    & 454 (0.1448\%)   & 0.1448 & 0.1483 \\
\bottomrule
\end{tabular}
\end{minipage}

\end{table}

Taken together, these findings show that SATD in SSW is characterized by both a low likelihood of removal and long remediation delays. In addition, projects with higher resolution rates do not consistently exhibit shorter removal times, suggesting that SATD removability and persistence capture distinct aspects of SATD behavior in SSW.
\vspace{-3mm}
\subsubsection{Comparison with OSS}
Prior SATD studies on OSS report
higher SATD removal rates and marked shorter persistence times. Potdar and Shihab~\cite{potdar2014exploratory} observed SATD resolution rates between 39.9\% and 74.6\% across three OSS projects 
, all 
exhibited low code ownership density (mean 2.54 and median of 2.33 KLOC per contributor). 
Bavota and Russo~\cite{bavota2016large} conducted an 
analysis on 159 projects (mean 4.52 KLOC per contributor) finding that approximately 57\% of SATD instances are eventually removed.
Maldonado et al.~\cite{maldonado2017empirical} analyzed five OSS projects (mean 4.00 and median of 2.77 KLOC per contributor) finding that between 40.5\% and 90.6\% 
of identified SATD comments get removed.  
Maldonando et al.~\cite{maldonado2017empirical} also analyzed the removal times of SATD in OSS. They found that in general SATD removal varies from project to project with medians ranging between 18.2–172.8 days and averages ranging between 82–613.2 days.  
Compared to OSS, SATD comments in SSW are both less likely to be removed and longer lived. Prior OSS studies consistently report that a majority of SATD is resolved within weeks or months. In contrast, our SSW mean and median removal times exceed the ranges of median and average values reported in OSS studies. This difference holds despite substantial variation in project size and contributor distribution across both domains.

To examine the role of project scale, we analyze the relationship between KLOC per contributor and SATD resolution behavior. Spearman rank correlation reveals a strong negative association between KLOC per contributor and SATD resolution rate ($\rho = -0.78$, $p = 0.013$), indicating that projects with fewer lines of code per contributor tend to remove a higher proportion of SATD comments. In contrast, the association between KLOC per contributor and removal time is weak and not statistically significant ($\rho = -0.17$, $p = 0.67$), indicating no meaningful monotonic relationship between project scale and the duration that SATD persists once introduced.
Importantly, even SSW projects with relatively low KLOC per contributor approaching or overlapping with the ranges reported for OSS, still exhibit lower SATD resolution rates and longer removal times than those observed in OSS. This suggests that the persistence of SATD in SSW cannot be explained by code ownership density or project scale alone. Instead, these findings point to domain specific factors in SSW collectively contributing to the unusually persistent nature of SATD in SSW.
\vspace{-3mm}
\vspace{-1mm}
\subsection{RQ3: To what extent does SATD propagate across SSW artifacts?}
\vspace{-1mm}
To examine how SATD propagates across SSW artifacts, we performed an exhaustive traversal of artifact linkage chains up to a depth of four, originating from comments, commits, pull requests, and issues. For each traversal, we tracked whether SATD persisted across linked artifacts and analyzed associated priority scores along continuous SATD chains.

Table~\ref{tab:satd_chain_length_combined} shows that SATD propagation across multiple artifacts is structurally possible but uncommon. 
The majority of SATD is contained to isolated artifacts as represented by length one (3.86\%), indicating that most SATD instances are isolated rather than carried forward across development artifacts. SATD only chains of length greater than or equal to two, where all links in the chain contain SATD, are rare, occurring at less than 0.2\%.
Despite their rarity, longer SATD chains exhibit higher mean and median priority scores than isolated SATD instances. 
This suggests that when SATD persists across multiple artifacts it is generally associated with textual content that suggests higher urgency. 
When tracing SATD originating from code comments (Table \ref{tab:satd-propagation}), over half of instances (52.38\%) remain confined to the comment itself, indicating that many comment level SATD expressions capture localized debt that does not explicitly influence downstream development artifacts. However, a substantial proportion is traceable in commits (40.87\%), suggesting that comment level SATD frequently informs immediate code changes or implementation decisions.
\setlength{\textfloatsep}{6pt plus 2pt minus 2pt}
\begin{table}
\captionsetup{skip=0pt}
\centering
\footnotesize
\setlength{\tabcolsep}{4pt}

\begin{minipage}[c]{0.48\textwidth} 
\centering
\caption{SATD propagation depth across artifacts (ordered path from comments)}
\label{tab:satd-propagation}
\begin{tabular}{lrr}
\toprule
\textbf{Propagation Path} & \textbf{Count} & \textbf{Percentage (\%)} \\
\midrule
comment only                 & 9,927  & 52.38 \\
comment $\rightarrow$ commit  & 7,746  & 40.87 \\
comment $\rightarrow$ commit $\rightarrow$ PR & 1,206  & 6.36 \\
comment $\rightarrow$ commit $\rightarrow$ PR $\rightarrow$ issue & 74 & 0.39 \\
\bottomrule
\end{tabular}
\end{minipage}
\hfill
\begin{minipage}[c]{0.48\textwidth}  
\centering
\caption{SATD propagation depth across artifacts (ordered path from issues)}
\label{tab:satd-propagation-inverted}
\begin{tabular}{lrr}
\toprule
\textbf{Propagation Path} & \textbf{Count} & \textbf{Percentage (\%)} \\
\midrule
issue only                         & 4,000  & 78.68 \\
issue $\rightarrow$ PR              & 1,023  & 20.12 \\
issue $\rightarrow$ PR $\rightarrow$ commit & 44  & 0.87 \\
issue $\rightarrow$ PR $\rightarrow$ commit $\rightarrow$ comment & 17  & 0.33 \\
\bottomrule
\end{tabular}
\end{minipage}

\end{table}
Propagation beyond commits becomes increasingly rare, with only 6.36\% reaching PRs and less than 1\% extending to issues. This drop off indicates that comment level SATD primarily influences short term development actions rather than long term planning, consistent with the role of comments as localized expressions of technical compromise.

In contrast, SATD originating from issues (Table \ref{tab:satd-propagation-inverted}) exhibits even more limited propagation. Most issue level SATD remains isolated within the issue itself (78.68\%), with only a minority propagating to PRs (20.12\%) and very few extending to commits or comments. This pattern suggests that SATD expressed in issues often reflects high-level concerns, exploratory discussion, or deferred ideas that do not translate directly into concrete code-level debt acknowledgments.

Taken together, these propagation patterns reveal an asymmetric flow of SATD across artifacts. SATD expressed closer to the codebase is more likely to influence subsequent development actions, whereas SATD documented in higher-level planning artifacts rarely propagates downward into implementation-level artifacts. This asymmetry highlights a disconnect between strategic debt discussions and operational development activities in SSW projects.
\vspace{-2mm}

\subsection{RQ4: How does PR and issue length relate to SATD occurrence, sentiment, and priority in SSW?}
\vspace{-1mm}
To understand how artifact length influences SATD in SSW, we examined the relationship between artifact length (measured in tokens) and SATD rates, sentiment scores, and priority scores. 
We filtered artifacts to include only those with at least 10 tokens and divided both issues and PRs into 10 quantile based bins for length analysis.

\begin{figure}
\centering

\begin{minipage}[c]{0.45\textwidth} 
    \centering
    \includegraphics[width=\linewidth]{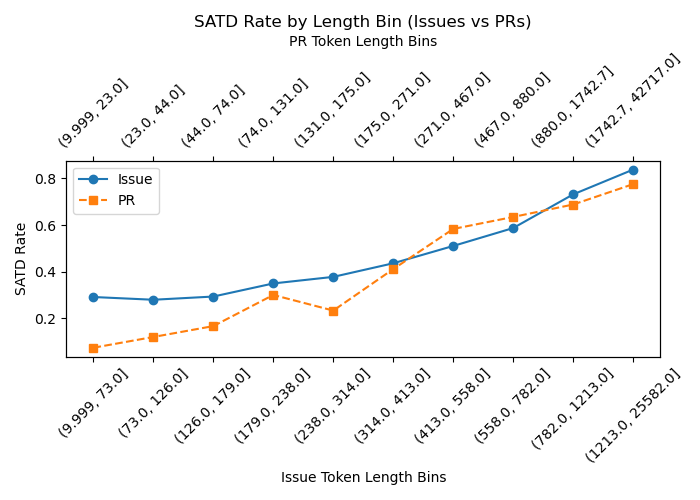}
     \caption{SATD rate by token length bin for issues and PRs.}
    \label{fig:satd_rate_length}
\end{minipage}
\hfill
\begin{minipage}[c]{0.45\textwidth}  
    \centering
    \includegraphics[width=\linewidth]{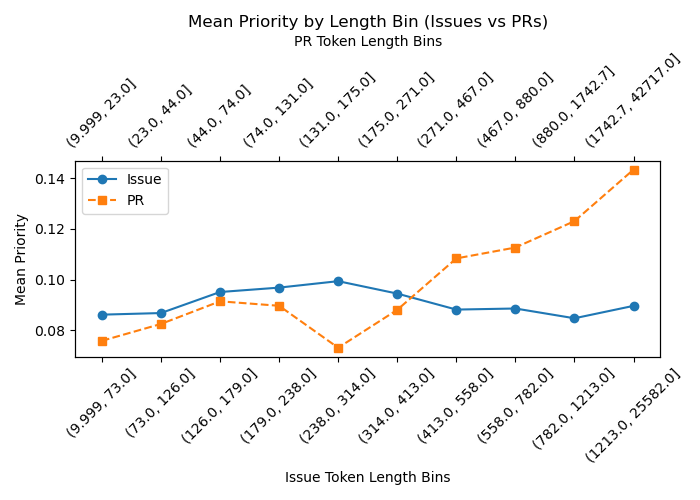}
     \caption{Mean SATD priority by token length bin for issues and PRs.}
    \label{fig:priority_length}
\end{minipage}
\hfill
\begin{minipage}[c]{0.45\textwidth} 
    \centering
    \includegraphics[width=\columnwidth]{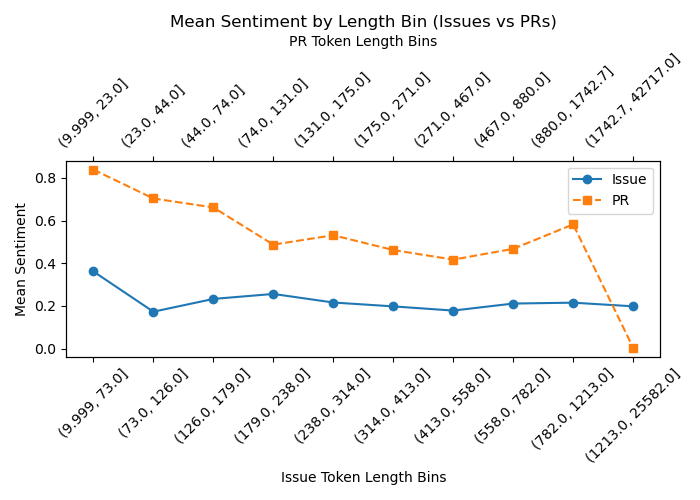}
    \caption{Mean sentiment score by token length bin for issues and PRs.}
    \label{fig:sentiment_length}
\end{minipage}\end{figure}

Our analysis revealed a strong positive relationship between artifact length and SATD occurrence for both issues and PRs (Figure \ref{fig:satd_rate_length}). For issues, SATD rates increased monotonically from 29.2\% in the shortest bin (10-73 tokens) to 83.8\% in the longest bin (1,213-25,582 tokens). PRs exhibited a similar but more pronounced trend, with SATD rates climbing from 7.4\% in the shortest bin (10-23 tokens) to 77.6\% in the longest bin (1,743-42,717 tokens). Notably, PRs often showed lower SATD rates than issues across corresponding length percentiles, particularly in the shorter bins where the difference was most pronounced. However, the gap narrowed substantially at higher lengths, with both artifact types converging around 78-84\% SATD rates in their longest bins. This suggests that while issues are more prone to containing technical debt discussions at shorter lengths, artifact complexity, as proxied by length becomes the dominant factor for SATD occurrence in both artifact types at higher token counts.

 \vspace{-1mm}

Priority scores showed distinct patterns across artifact types (Figure \ref{fig:priority_length}). For issues, mean priority remained relatively stable across all length bins, fluctuating narrowly between 0.085 and 0.099 with no clear monotonic trend. The slight variation observed suggests that priority in issues is largely independent of artifact length. PRs, however, exhibited a general upward trend in priority scores as length increased, rising from 0.076 in the shortest bin to 0.143 in the longest bin, an increase of approximately 88\%. This suggests that longer PRs tend to address higher priority SATD, while priority in issues is less dependent on artifact length. The increasing priority in longer PRs may reflect the fact that complex code changes often address critical architectural or design concerns that require extensive documentation, explanation, and justification, thereby generating longer artifact descriptions and discussions.

The relationship between artifact length and sentiment scores differed markedly between issues and PRs (Figure \ref{fig:sentiment_length}). Issues maintained relatively stable sentiment scores across all length bins, ranging from 0.17 to 0.36, with a slight declining trend in longer artifacts. The highest sentiment score for issues (0.36) occurred in the shortest bin, dropping to approximately 0.20 for the longest artifacts. In contrast, PRs demonstrated substantially higher initial sentiment scores (0.84 in the shortest bin) that declined sharply and progressively to near-zero (0.004) in the longest bin. This dramatic divergence suggests that longer PRs tend to discuss more technical or negative concerns, while issues maintain a more consistently neutral-to-slightly-positive emotional tone regardless of length. The high initial sentiment in short PRs may reflect straightforward, positive contributions such as simple bug fixes or minor enhancements, while complex PRs involve more critical technical discussions concerning architectural decisions or difficult implementation challenges.
\vspace{-3mm}

%% file: sec_discussion.tex
\vspace{-2mm}

This study provides the first multi-artifact, prioritization, sentiment, and propagation analysis of SATD in SSW, showing that it behaves differently from OSS SATD in management, and persistence.
Prioritization (RQ1): Debt close to code commits, comments, and pull requests receives higher priority than issue-level debt. Test and documentation debt dominate high priority artifacts, while scientific debt is lower, highlighting a mismatch between conventional heuristics and SSW needs. Negative sentiment amplifies perceived priority, confirming its value as a signal for intervention.
Resolution and persistence (RQ2): SATD in SSW is rarely removed, often persisting for long periods of time. Unlike OSS, where most debt resolves within weeks to months, the persistent nature of SATD in SSW may instead be a result of the demand to rapidly disseminate findings, or pressures stemming from evolving scientific requirements, consistent with prior findings that SSW evolves under distinct pressures~\cite{carver2016software}.
Propagation (RQ3): Most SATD remains localized, but rare cross artifact chains carry higher priority. Debt in code level artifacts rarely affects planning, and issue level debt seldom propagates to implementation, revealing a disconnect between strategic planning and operational development. Multi-artifact propagation patterns can thus serve as indicators of high impact debt.
Artifact characteristics (RQ4): Longer PRs and issues are more likely to contain SATD, with PR length correlating with priority escalation and negative sentiment.
Overall, SSW SATD management must account for long lived, domain specific debt, leveraging artifact links and sentiment to prioritize interventions. Propagation aware, multi-artifact analyses complement static methods, supporting scalable monitoring in collaborative, long lived SSW projects. SATD in SSW is persistent, nuanced, and context dependent; addressing it effectively requires tailored tools and methods that integrate artifact linkage, developer sentiment, and domain specific prioritization to promote sustainable, reproducible, and high quality scientific computation.
\vspace{-3mm}

%% file: sec_threats.tex
\vspace{-2mm}
\textbf{Construct and Internal Validity.} SATD identification relies on automated classification and may be affected by ambiguity in developer language or model inference errors. We mitigate this risk by using a classifier fine-tuned for SSW and multi-artifact contexts. SATD prioritization is heuristic-based and depends on a predefined keyword list, which may not fully reflect project-specific or scientific priorities. Propagation analysis traces structural artifact links and does not guarantee that identical underlying debt concerns are expressed across artifacts.

\textbf{External and Conclusion Validity.} Our analysis focuses on nine SSW repositories and may not generalize to all SSW or OSS projects. Observed correlations do not imply strict causation, and unobserved factors such as funding cycles or developer expertise may influence results. We encourage replication across additional SSW projects.
By acknowledging these threats, we provide context for interpreting our findings and encourage replication and extension in diverse SSW contexts.
\vspace{-3mm}

%% file: sec_conclusion.tex
\vspace{-2mm}
This study presents the first multi-artifact analysis of SATD priority, sentiment, resolution, and linkage in SSW.
We observe that SATD prioritization in SSW varies systematically across artifact types and debt categories, with code-proximate artifacts (commits, comments, pull requests) receiving higher priority than issues, and Test and Documentation Debt consistently prioritized above Scientific Debt.
Negative sentiment strongly correlates with higher priority across contexts, reinforcing its value as a signal of perceived urgency.
SATD in SSW exhibits notable persistence compared to OSS, with only 37.8\% of instances removed and mean and median removal times of approximately 690 and 279 days, respectively. This indicates substantially longer SATD lifespans than those reported in OSS studies, which typically report median removal times ranging from 18.2 to 172.8 days and mean values from 82 to 613.2 days. In addition, reported OSS resolution rates vary widely across studies (approximately 40\%–75\%), whereas SSW resolution rates are lower in aggregate and vary between 15.0\% and 58.6\% across projects.
While SATD propagation across artifacts is possible, it remains uncommon, with most instances (78.68\% for issues, 52.38\% for comments) confined to their originating artifact.
When propagation occurs, higher associated priority scores indicate greater severity. Artifact length serves as a strong proxy for complexity and SATD presence, with longer artifacts more likely to contain SATD; however, issues and pull requests exhibit distinct patterns.
These findings reveal insights into SATD management in SSW and fundamental differences from OSS practices, characterized by exceptional persistence and low removal rates.
The results suggest that SATD management and prioritization approaches developed for OSS may require adaptation to address SSW’s unique characteristics, particularly the domain-specific nature of Scientific Debt and persistent unresolved compromises.
\vspace{-2mm}

%% file: sec_acknowledgments.tex
\vspace{-2mm}
This material is based upon work supported by the U.S. Department of Energy, Office of Science, Office of Workforce Development for
Teachers and Scientists, Office of Science Graduate Student Research (SCGSR) program. The SCGSR program is administered by the
Oak Ridge Institute for Science and Education (ORISE) for the DOE. ORISE is managed by ORAU under contract number
DESC0014664. We acknowledge the support of the Consortium for Open-Source
Research Software Advancement (CORSA), a project supported
by the U.S. Department of Energy, Office of Science, Office of
Advanced Scientific Computing Research, Next-Generation Scientific Software Technologies program, under contract number
DE-AC05-00OR22725. All opinions expressed in this paper are the author’s and do not necessarily reflect the policies and views of DOE,
ORAU, or ORISE.
\vspace{-1mm}